\documentclass[twocolumn, tight]{aastex62}

\usepackage{natbib}
\usepackage{amssymb, amsmath}
\usepackage{appendix}
\usepackage{graphicx}
\usepackage{enumerate}% http://ctan.org/pkg/enumerate
\usepackage{xcolor}
\usepackage{ulem}

\received{December 2018}
\revised{May 2019}
\accepted{May 2019}
\shorttitle{Extreme Line Ratios in ULIRGS}
\shortauthors{Brown \& Wilson}
\begin{document}

% Further Evidence for a Top-Heavy IMF in Nearby ULIRGs?
% Unusual Isotopologue Line Ratios in ULIRGS: Evidence of a top-heavy IMF?
\title{Extreme CO Isotopologue Line Ratios in ULIRGS: Evidence for a top-heavy IMF}

\author[0000-0003-1845-0934]{Toby~Brown*}
\email{tobiashenrybrown@gmail.com}
\affil{Department of Physics and Astronomy, McMaster University,
     1280 Main St. W., Hamilton, Ontario L8S 4M1, Canada}

\author[0000-0001-5817-0991]{Christine~D.~Wilson}
\affiliation{Department of Physics and Astronomy, McMaster University,
               1280 Main St. W., Hamilton, Ontario L8S 4M1, Canada}



\begin{abstract}
We present high-resolution ALMA observations of the C$^{18}$O, $^{13}$CO and $^{12}$CO $J$=1-0 isotopologues in 3 nearby ultra-luminous infrared galaxies (ULIRGS; Arp 220, IRAS 13120-5453, IRAS 17208-0014) and 1 nearby post-merger galaxy (NGC 2623). In all 4 systems, we measure high $^{12}$CO/C$^{18}$O and $^{12}$CO/$^{13}$CO integrated line ratios while the $^{13}$CO/C$^{18}$O ratio is observed to be extremely low in comparison to typical star-forming disks, supporting previous work. We investigate whether these unusual line ratios are due to dynamical effects, astrochemistry within the gas, or nucleosynthesis in stars. Assuming both lines are optically thin, low $^{13}$CO/C$^{18}$O values suggest that C$^{18}$O is more abundant than $^{13}$CO in the interstellar medium of these systems. A plausible explanation is that local ULIRGs and their progeny have an excess in massive star formation; in other words, they are producing a top-heavy stellar initial mass function.
\end{abstract}


\section{Introduction}
\label{introduction}
It is well established that the merging of two galaxies has a significant effect on the system's interstellar medium (ISM) and star formation properties. Radial atomic and molecular gas flows towards the merger's center dissipate energy and angular momentum \citep{Barnes1991}. Once there, increased densities and dust contents aid the conversion of atomic gas into its molecular, star-forming phase, and strong tidal forces combine with frequent molecular cloud collisions to elevate star formation rates \citep[SFRs;][]{Mirabel1989,Sanders1996,Genzel1998}.  In extreme cases, where the merging pair is both massive and gas-rich, SFRs can reach hundreds or even thousands of solar masses per year, heating the interstellar dust such that infrared luminosities exceed 10$^{12}$ L$_\odot$ \citep[e.g.,][]{Armus1989,Barnes1992,Murphy1999}. In the local Universe, these objects are often known as ultra-luminous infrared galaxies (ULIRGs).

While ULIRGs are relatively rare in comparison to typical nearby galaxies, extreme star-forming systems become many hundreds of times more common at high redshift, where the bulk of galaxy formation took place \citep[see review by][and references therein]{Lonsdale2006}. Indeed, these so-called starburst galaxies are often cited as a major driver of the peak in the cosmic SFR density at $z\sim 2-3$ \citep{Madau2014}. The proximity and luminosity of nearby ULIRGs enable highly detailed, resolved observations and their extreme SFRs make them compelling analogues of high-$z$ galaxies, ideal for testing current theories of star formation and galaxy evolution.

Regardless of the galaxy and its redshift, the determination of many intrinsic galaxy properties (including SFR and stellar mass) must assume an initial distribution of birth masses for stars, the so-called stellar initial mass function (IMF). The precise form of the stellar IMF plays a leading role in current star formation theories; however, the past three decades have seen considerable debate within the literature on the shape and variable nature of the IMF \citep[see reviews by][]{Chabrier2003,Bastian2010}. Historically, IMF estimates have relied on infrared, optical, and ultraviolet observations of either individual stars or integrated stellar populations. However, the large quantities of interstellar dust present in ULIRGs heavily obscure light at these wavelengths; thus, traditional techniques for measuring the IMF are not possible for starburst galaxies at any redshift.

Less traditional methods using submillimeter observations of the molecular gas content that are immune to dust attenuation hold much promise as sensitive probes of the IMF. In particular, there is a considerable body of work that has had some success using the measurement of isotopologue (isotope bearing molecule) line ratios in the ISM \citep[e.g.,][]{Sage1991, Henkel1993, Papadopoulos1996, Meier2004, Danielson2013, Sliwa2017, Zhang2018}. This approach relies upon differences in stellar evolutionary paths which mean the production of carbon, oxygen, and their isotopes varies with individual star mass. High-mass stars ($\gtrsim 8 \, {\rm M}_\odot$) are the predominant source of the oxygen-18 isotope ($^{18}$O) while the carbon-13 isotope ($^{13}$C), a by-product of the CNO-cycle, is primarily synthesized in low- to intermediate-mass stars \citep[$< 8 \, {\rm M}_\odot$;][]{Iben1975, Wilson1992, Meynet2002, Matteucci2012}. 
% Second, the variation in timescales required to synthesize elements and eject them into the ISM means that $^{12}$C and $^{16}$O are released at faster rates than $^{13}$C and $^{18}$O respectively \citep{Wilson1992, Romano2017}.
Assuming that the isotopologue line ratio $^{13}$CO/C$^{18}$O traces the abundance ratio of $^{13}$C/$^{18}$O in the high-pressure ISM of ULIRGs,  $^{13}$CO/C$^{18}$O can be used as chemical tracer of the stellar IMF and the star formation history \citep[for an extensive discussion of this approach see Section 2.3 of][]{Romano2017}.

In the Milky Way and nearby normal star-forming galaxies, $^{13}$CO is typically 7-10 times brighter than C$^{18}$O \citep{Langer1993,Jimenez-Donaire2017}. However, much lower $^{13}$CO/C$^{18}$O values have been observed in ULIRGs and starbursting systems. Both \citet{Greve2009} and \citet{Matsushita2009} find $^{13}$CO/C$^{18}$O to be around unity or below in the ULIRG Arp 220, with the latter suggesting that either optical depth effects or Arp 220's ongoing star formation activity are responsible for the observed strength of C$^{18}$O with respect to $^{13}$CO. More recently, \citet{Sliwa2017} observed similarly low $^{13}$CO/C$^{18}$O values in another ULIRG, IRAS 13120-5453, along with an unusually high $^{12}$CO/$^{13}$CO ratio of $>60$. The authors find these unusual line ratios to be consistent with a very recent star formation episode ($<7$ Myr) and/or a top-heavy IMF. An IMF biased in favor of massive stars has been suggested previously by \citet{Sage1991}, \citet{Henkel1993}, and \citet{Papadopoulos1996} to explain an observed overabundance of $^{18}$O in the starburst nuclei of 4 nearby active galaxies. \citet{Meier2004} also attribute observed enhancements in $^{12}$CO and C$^{18}$O relative to the $^{13}$CO in the nucleus of the nearby barred spiral galaxy NGC 6946 to the same effect. At redshift $z\sim 2-3$, the top-heavy IMF scenario has been invoked to explain low $^{13}$CO/C$^{18}$O ratios found in a small number of gravitationally lensed starbursts \citep{Danielson2013,Zhang2018}.

The challenge now is to understand whether the unusual CO isotopologue line ratios observed in extreme star-forming galaxies can be attributed primarily to their recent star formation activity or to a more fundamental difference in the IMF of these systems, or, as is possible, neither.

Building on previous work, this paper aims to investigate the prevalence, robustness, and physical cause of unusual CO isotopologue intensity ratios in systems that span the merger sequence. To do so, we combine new and archival Atacama Large Millimeter/submillimeter Array (ALMA) spatially resolved observations of $^{12}$CO, $^{13}$CO and C$^{18}$O in dense molecular gas for a sample of 3 nearby ULIRGs and 1 post-merger galaxy. In Section \ref{sec:observations}, we describe the ALMA data reduction and present our observations. Section \ref{sec:LineRatios} discusses several possible explanations for our results, including optical depth effects, selective photodissociation, a recent burst of star formation, and an excess in massive star formation. Section \ref{sec:conclusions} presents our conclusions. All distant dependent quantities are derived assuming a $\Lambda$CDM cosmology with $\Omega = 0.3$, $\Lambda = 0.7$, and a Hubble constant $\text{H}_{0} = 70\; \text{km s}^{-1} \;\text{Mpc}^{-1}$.

\section{Observations}
\label{sec:observations}

\begin{figure*}
\centering
\includegraphics[width=\textwidth]{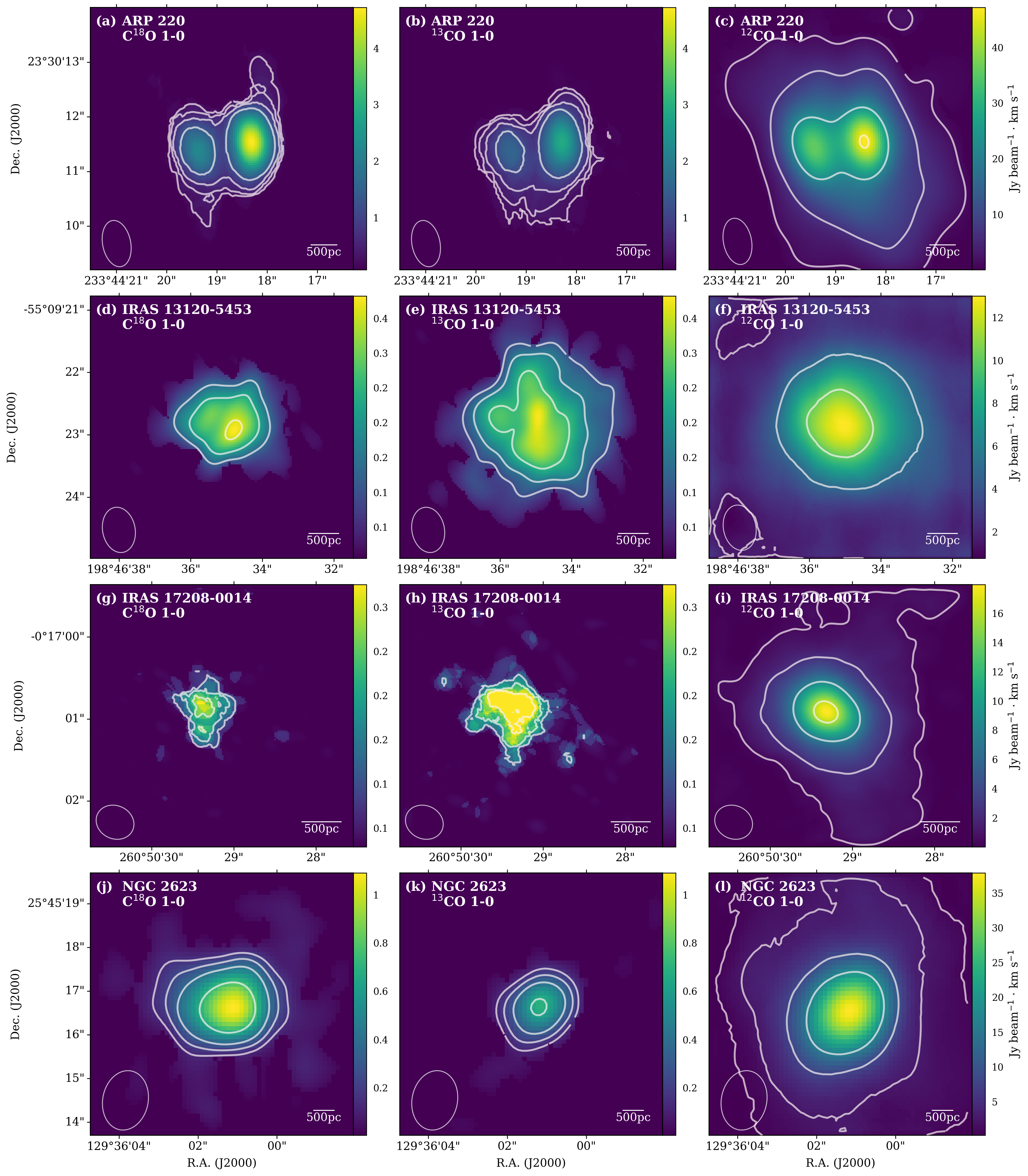}
\caption{Integrated intensity maps of C$^{18}$O $J$=1-0 (left column), $^{13}$CO $J$=1-0 (middle column) and $^{12}$CO $J$=1-0 (right column) for Arp 220 (a, b, c), IRAS 13120-5453 (d, e, f), IRAS 17208-0014 (g, h, i), and NGC 2623 (j, k, l). For scale, a 500pc physical size is depicted by the bar in the bottom right of each panel and the ALMA synthesized beam of each observation is shown by the ellipse in the bottom left. To facilitate comparison, we match the beam and pixel resolution of the line observations for each galaxy using CASA routines. The threshold for each map is 3 times the channel rms given in Table \ref{table:properties}. Contours are set to 3$\sigma$, 5$\sigma$,  8$\sigma$ and 15$\sigma$ for C$^{18}$O and $^{13}$CO, and 3$\sigma$, 15$\sigma$, 50$\sigma$ and 100$\sigma$ for $^{12}$CO maps. Here $\sigma = {\rm rms} \cdot \sqrt{\Delta V_{chan} \cdot V_{line}}$, where $\Delta V_{chan}$ is the channel width in km s$^{-1}$ and $V_{line}$ is the velocity extent over which the line is integrated to make the moment map.}
\label{fig:momentMaps}
\end{figure*}

\begin{figure*}
\centering
\includegraphics[width=\textwidth]{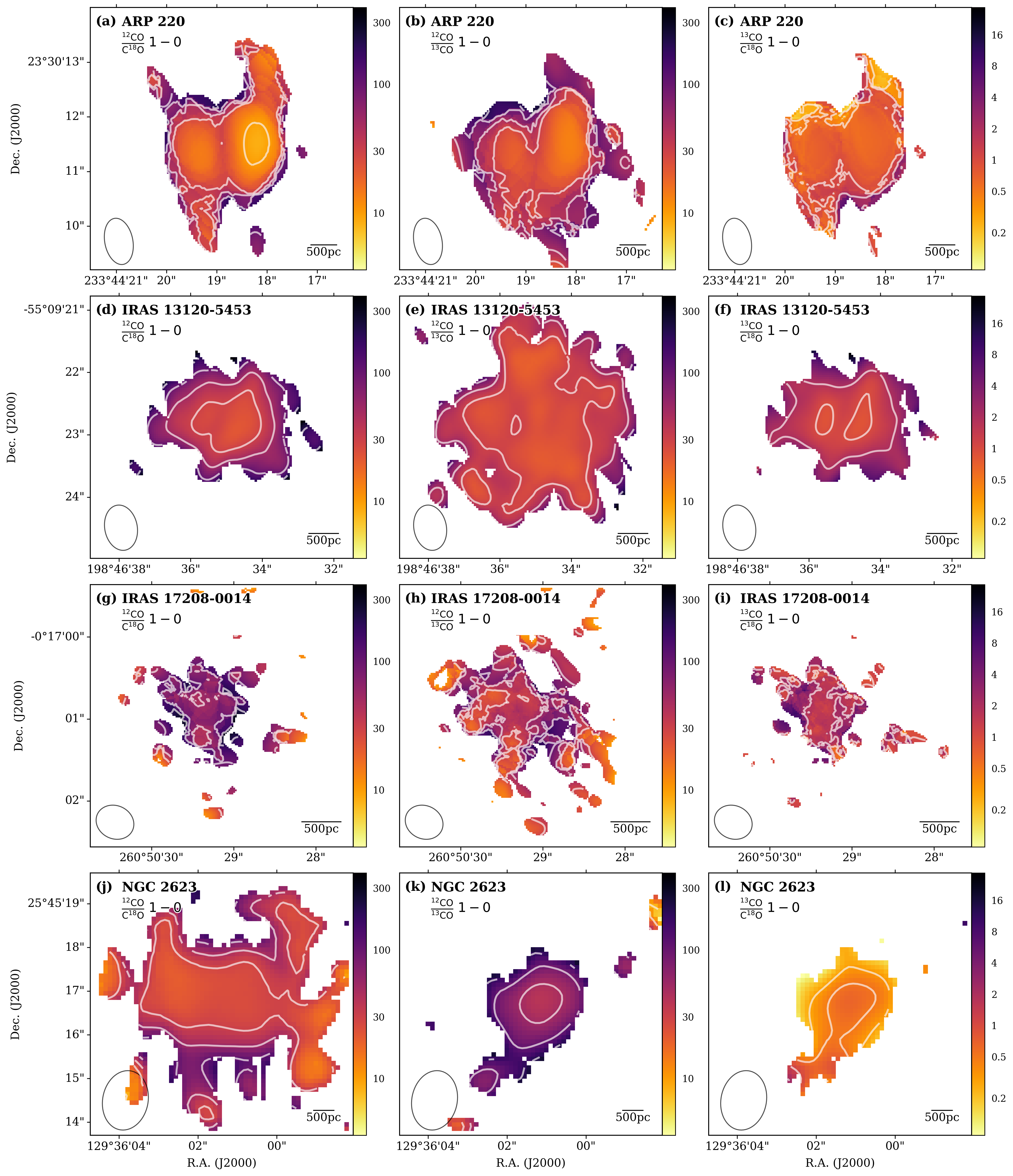}
\caption{$^{12}$CO/C$^{18}$O $J$=1-0 (left column), $^{12}$CO/$^{13}$CO $J$=1-0 (middle column) and $^{13}$CO/C$^{18}$O $J$=1-0 (right column) line ratio maps for Arp 220 (a, b, c), IRAS 13120-5453 (d, e, f), IRAS 17208-0014 (g, h, i), and NGC 2623 (j, k, l). The ALMA synthesized beam size is shown by the eclipse in the bottom left while a 500pc scale bar is given in the bottom right. Color maps are shown in log-scale. For $^{12}$CO/C$^{18}$O and  $^{12}$CO/$^{13}$CO line ratio maps, contour values are 10, 30, 50, 100 and 200, while for $^{13}$CO/C$^{18}$O maps they are 0.3, 0.5, 1 and 2.}
\label{fig:ratioMaps}
\end{figure*}

This paper combines new and archival ALMA 12m-array observations of C$^{18}$O, $^{13}$CO and $^{12}$CO $J$=1-0 lines in 4 nearby ($<$200 Mpc) advanced mergers: Arp 220, IRAS 13120-5453, IRAS 17208-0014, and NGC 2623. With infrared luminosities, L$_{\rm IR}$, exceeding 10$^{12}$ L$_{\odot}$, Arp 220, IRAS 13120-5453, and IRAS 17208-0014 are all ULIRGs, while NGC 2623 is a post-merger galaxy with L$_{\rm IR}$ = 10$^{11.6}$ L$_{\odot}$ \citep{Armus2009}. The targets, their distances, infrared (IR) luminosities, and individual isotopologue line properties are listed in Table \ref{table:properties}. The combination of new and archival data means that our targets were observed across ALMA cycles 2-5. Each calibrated measurement set was reduced using the appropriate \textit{Common Astronomy Software Applications} \citep[CASA;][]{McMullin2007} version recommended by ALMA's \href{https://almascience.nrao.edu/processing/science-pipeline#version}{science pipeline documentation}. The calibrated measurement sets are continuum subtracted and reduced using CASA v5.1.1. We image all data using CASA's \texttt{TCLEAN} command with a {\em Briggs} weighting \citep{Briggs1995}, robust parameter of 0.5, and rms threshold of 2$\sigma$. We set the image channel width to 20 km s$^{-1}$ for $^{12}$CO and 35 km s$^{-1}$ for C$^{18}$O and $^{13}$CO.

\startlongtable
\begin{deluxetable*}{lcccc}
\tabletypesize{\footnotesize}
\tablewidth{0pt}
\small
\tablecaption{List of target galaxies and their C$^{18}$O $J$=1-0, $^{13}$CO $J$=1-0 and $^{12}$CO $J$=1-0 line and ratio properties. The associated measurement errors are quoted (no brackets) and ALMA band 6 calibration uncertainties are provided in brackets where} appropriate. \label{table:properties}
\tablehead{\colhead{Galaxy}               &
           \colhead{Arp 220}              &
           \colhead{IRAS 13120-5453}      &
           \colhead{IRAS 17208-0014}      &
           \colhead{NGC 2623}}

\startdata
D$_L$ [Mpc] 											 & 79                             & 135                            & 189                          & 80 \\
log $L_{IR}$ [$L_\odot$] 								 & 12.2                           & 12.3                           & 12.4                         & 11.4 \\
Resolution ["] 											 & 0.86  $\times$ 0.51            & 0.73 $\times$ 0.52             & 0.47 $\times$ 0.40           & 1.37 $\times$ 1.04 \\
C$^{18}$O $J$=1-0 flux [Jy km s$^{-1}$] 				 & 10   $\pm$ 0.4 [$\pm$ 0.7]    & 1   $\pm$ 0.2 [$\pm$ 0.1]    & 0.5  $\pm$ 0.1 [$\pm$ 0.04]  & 2   $\pm$ 0.2 [$\pm$ 0.1] \\
$^{13}$CO $J$=1-0 flux [Jy km s$^{-1}$] 				 & 8   $\pm$ 0.3 [$\pm$ 0.5]    & 2   $\pm$ 0.2 [$\pm$ 0.2]    & 1  $\pm$ 0.2 [$\pm$ 0.1]   & 0.6   $\pm$ 0.1 [$\pm$ 0.04] \\
$^{12}$CO $J$=1-0 flux [Jy km s$^{-1}$] 				 & 294 $\pm$ 3 [$\pm$ 21]   & 156 $\pm$ 0.8 [$\pm$ 11]   & 70 $\pm$ 1 [$\pm$ 5]   & 109 $\pm$ 1 [$\pm$ 8] \\
C$^{18}$O $J$=1-0 rms [mJy beam$^{-1}$] (35 km s$^{-1}$) & 0.7                            & 0.3                            & 0.3                          & 0.4 \\
$^{13}$CO $J$=1-0 rms [mJy beam$^{-1}$] (35 km s$^{-1}$) & 0.7                            & 0.3                            & 0.3                          & 0.4 \\
$^{12}$CO $J$=1-0 rms [mJy beam$^{-1}$] (20 km s$^{-1}$) & 4.2                            & 1.1                            & 1.4                          & 1.9 \\
C$^{18}$O V$_{\rm min}$:V$_{\rm max}$ [km s$^{-1}$] 	 & 5130:5550                      & 8840:9300                      & 12040:12320                  & 4090:4510 \\
$^{13}$CO V$_{\rm min}$:V$_{\rm max}$ [km s$^{-1}$] 	 & 4010:4320                      & 7760:8140                      & 10990:11510                  & 5240:5590 \\
$^{12}$CO V$_{\rm min}$:V$_{\rm max}$ [km s$^{-1}$] 	 & 5060:5660                      & 8790:9240                      & 1970:12650                   & 5240:5660 \\
Mean $^{12}$CO/C$^{18}$O $J$=1-0 		                 & 39 $\pm$ 2 [$\pm$ 0.1] & 63  $\pm$ 4 [$\pm$ 0.2]   & 81 $\pm$ 6 [$\pm$ 0.2]   & 40 $\pm$ 2 [$\pm$ 0.1] \\
Mean $^{12}$CO/$^{13}$CO $J$=1-0 		                 & 48 $\pm$ 2 [$\pm$ 0.1] & 36  $\pm$ 0.8 [$\pm$ 0.1] & 45 $\pm$ 2 [$\pm$ 0.1]   & 92 $\pm$ 10 [$\pm$ 0.5] \\
Mean $^{13}$CO/C$^{18}$O $J$=1-0 		                 & 0.9  $\pm$ 0.05        & 2   $\pm$ 0.2             & 2  $\pm$ 0.2             & 0.5 $\pm$ 0.1  \\
Median $^{12}$CO/C$^{18}$O $J$=1-0 						 & 26 $\pm$ 3 [$\pm$ 3] & 48  $\pm$ 5  [$\pm$ 3]      & 70 $\pm$ 8 [$\pm$ 7]     & 31 $\pm$ 2 [$\pm$ 3]   \\
Median $^{12}$CO/$^{13}$CO $J$=1-0 						 & 37 $\pm$ 2 [$\pm$ 4] & 30  $\pm$ 1  [$\pm$ 2]      & 38 $\pm$ 2 [$\pm$ 4]     & 81 $\pm$ 13 [$\pm$ 6]   \\
Median $^{13}$CO/C$^{18}$O $J$=1-0 						 & 0.7  $\pm$ 0.06 	   & 2   $\pm$ 0.2             & 2  $\pm$ 0.2             & 0.4 $\pm$ 0.2 \\
\enddata
\end{deluxetable*}

For each galaxy, we match the pixel size, beam resolution and length of the minimum baseline for all 3 emission lines using \texttt{TCLEAN}'s \texttt{CELL}, \texttt{UVTAPER} and \texttt{UVRANGE} parameters, with the final beam matching achieved using \texttt{IMSMOOTH}.

Figure \ref{fig:momentMaps} shows $^{12}$CO, C$^{18}$O and $^{13}$CO $J$=1-0 integrated intensity maps for the 4 galaxies in our sample.  All moment maps have a 3$\sigma$ threshold and are primary beam corrected using the CASA routine \texttt{IMPBCOR}. The beam sizes are illustrated by the ellipse in the lower left corner and a 500pc scale bar is provided for reference. We note that IRAS 13120-5453's distribution of $^{13}$CO emission in panel (Fig. \ref{fig:momentMaps}e) differs from that previously reported by \citet{Sliwa2017}, who found a central hole in the emission. We do not recover the central depression in emission found by \citet{Sliwa2017} in either their data (resolution = 0.55" $\times$ 0.41") or the more recent higher resolution data (0.4" $\times$ 0.26") with which it is combined in this work. We suggest that this may be due to the manual calibration conducted by those authors but cannot confirm this. The C$^{18}$O distribution for IRAS 13120-5453 and IRAS 17208-0014 is less extended than $^{13}$CO which is in turn more compact than $^{12}$CO. Interestingly, this is not the case for Arp 200 and NGC 2623. In the former, C$^{18}$O and $^{13}$CO are distributed over the same area while in the latter, C$^{18}$O is more extended throughout the disk than $^{13}$CO.

The double nuclei of the merger system, Arp 220, are visible in all 3 emission lines while, despite having undergone a major merger at some point in their past, IRAS 13120-5453, IRAS 17208-0014, and NGC 2623 all distribute their molecular gas over a single nucleus. Combined with their different infrared luminosities, these morphological differences suggest our sample can be considered snapshots of different stages of the merger sequence, evolving from the bright ongoing merger (Arp 220), through the late stages of merging (IRAS 13120-5453, IRAS 17208-0014) and, finally, resulting in the fainter post-merger system (NGC 2623).

Figure \ref{fig:ratioMaps} shows the intensity ratio maps of three line ratios for each target. The ratios shown are $\frac{\rm   ^{12}CO}{\rm C^{18}O} \, J=1-0$, $\frac{\rm ^{12}CO}{\rm ^{13}CO} \, J=1-0$, and $ \frac{\rm ^{13}CO}{\rm C^{18}O} \, J=1-0$. To make these ratio maps, we convert each moment map in Figure \ref{fig:momentMaps} from Jy beam$^{-1}$ km s$^{-1}$ to K km s$^{-1}$ before computing the ratio using CASA's \texttt{IMMATH} routine. In all 4 targets, the values of $^{12}$CO/C$^{18}$O $J$=1-0 are consistent with those reported previously for Mrk 231, IRAS 13120-5453, and Arp 220 \citep[$\sim$20-140;][]{Gonzalez2014,Falstad2015,Sliwa2017}. Similarly, the $^{12}$CO/$^{13}$CO $J$=1-0 line ratios exhibit a large range of values, from $\sim20-30$ in central regions of Arp 220 and IRAS	13120-5453 to $\sim50 - 150$ across the disks of IRAS 17208-0014 and NGC 2623.

Excluding potentially unreliable measurements in the outer regions where signal-to-noise is low and emission regions are smaller than the beam, we observe $^{12}$CO/$^{13}$CO ratios over $\sim100$ in IRAS 17208-0014 and NGC 2623. In the case of IRAS 13120-5453, we do not recover a robust $^{12}$CO/$^{13}$CO measurement that agrees with the very high values ($>100$) reported by \citet{Sliwa2017}. This is due to the reduced central $^{13}$CO intensities recovered by those authors which we do not find in our analysis.

Most intriguingly, each system has C$^{18}$O emission that is relatively strong compared to $^{13}$CO, particularly across the central regions where mismatches in UV coverage and differences between the intrinsic size of emission regions are not likely to be influencing the ratios. The $^{13}$CO/C$^{18}$O $J$=1-0 line ratio dips well below the typical values found in normal star-forming galaxies \citep[$\sim$7-10; e.g.,][]{Jimenez-Donaire2017}. Indeed, there are large areas in each system where the line ratio is actually inverted and C$^{18}$O line emission is stronger than $^{13}$CO.

\section{The Origins of Extreme Isotopologue Line Ratios}
\label{sec:LineRatios}
We now discuss six potential explanations for the unusual isotopologue ratios presented in the previous section. The first scenario is dynamically driven changes in the molecular gas properties, the second is the optical depth of the lines, the next two are astro-chemical effects occurring within the gas, and the last two are the result of nucleosynthesis in stars.

\subsection{Radial Flows of Molecular Gas}
Dilution of abundances can occur as pristine gas flows from the outer regions of the merging system towards the center. The main problem with this scenario is that the $^{13}$CO/C$^{18}$O ratio is expected to increase with galactic radius  \citep{Wouterloot2008, Jimenez-Donaire2017}. In addition, Herschel studies of OH and H$_2$O have found $^{16}$O/$^{18}$O $>$500 for mergers where gas inflow is still strongly ongoing \citep[young mergers; e.g.,][]{Falstad2015,Konig2016} and $^{16}$O/$^{18}$O $<$150 for mergers with no strong evidence of inflowing gas \citep[advanced mergers; e.g.,][]{Gonzalez2014,Falstad2015}. Thus, we do not expect the C$^{18}$O abundance to increase relative to $^{13}$CO with gas inflow, yet we observe C$^{18}$O emission to be strong relative to $^{13}$CO in all 4 targets.
% Additionally, \citet{Rupke2008} find that dilution of central abundances due to inflow in ULIRGs is typically a factor of 2. Following this, a maximum $^{12}$CO/$^{13}$CO of $\sim60$ should be observed in ULIRGs.

\subsection{Optical Depth Effects}
One possibility is that self-absorption due to high molecular gas column densities, rather than changes in isotopologue abundance, is the root cause of the low $^{13}$CO/C$^{18}$O ratio. A scenario where either one or both of the $^{13}$CO and C$^{18}$O lines is optically thick is of particular concern in Arp 220 where high density gas tracers (e.g., HCN, HCO$^{\rm +}$) have been observed and, in the core of the Western nucleus, dust is known to be optically thick at $\lambda = 2.6 \,{\rm mm}$ with molecular hydrogen (H$_2$) column densities exceeding N$_{\rm H_2} \sim 10^{26} \, {\rm cm}^{-2}$ \citep{Greve2009,Scoville2017}.
% The $^{12}$CO/$^{13}$CO and $^{12}$CO/C$^{18}$O line ratios decrease towards the center of each disk, behaviour that is not expected for increasing optical depth of the $^{13}$CO and C$^{18}$O lines in the central regions.

When analyzing $^{13}$CO and C$^{18}$O lines in a sample of high-$z$ starburst galaxies, \citet{Zhang2018} find that H$_2$ column densities of N$_{\rm H_2} \sim 10^{25} \, {\rm cm}^{-2} \, {\rm or} \, \sim 10^{26} \, {\rm cm}^{-2}$ are required for $^{13}$CO/C$^{18}$O to approach unity in local thermodynamic equilibrium (LTE) or non-LTE conditions, respectively. While this is consistent with the highest column densities found in the center of Arp 220's Western nucleus, it remains an order of magnitude or more greater than the H$_2$ column densities of Arp 220's larger disk structures, as well as the molecular gas in IRAS 13120-5453, IRAS 17208-0014, and NGC 2623 \citep[N$_{\rm H_2} \sim 10^{21} - 10^{24}\, {\rm cm}^{-2}$;][]{Casoli1988,Aalto2015,Privon2017,Scoville2017}. \citet{Zhang2018} also find that even moderate $^{13}$CO optical depths ($\tau_{^{13}{\rm CO}}\sim0.2-0.5$) do not cause the $^{13}$CO/C$^{18}$O line ratio to deviate significantly from more typical values ($^{13}$CO/C$^{18}$O $\sim 7$). We note that the lowest average $^{13}$CO/C$^{18}$O value occurs in NGC 2623, a system where $^{13}$CO is unlikely to be optically thick.

Thus, although we cannot completely exclude optical depth effects, especially in Arp 220, we find it unlikely that this is the primary cause of the unusual ratios seen across the disks of every galaxy in our sample. This conclusion is supported by \citet{Martin2019} who account for optical depths in their analysis of CO isotopologues across the central regions of the nearby starburst galaxy NGC 253. However, to pro\textbf{}perly test this assumption, observations of multiple line transitions are required to measure excitation conditions and derive the optical depths. This will be the focus of future work. 

\subsection{Preferential Photodissociation}
The low $^{13}$CO/C$^{18}$O ratios also rule out preferential photodissociation of $^{13}$CO molecules compared to C$^{18}$O, especially in the inner regions. This is because C$^{18}$O would also undergo preferential photodissociation and at a faster rate than $^{13}$CO \citep{vanDishoeck1988,Casoli1992}; thus, preferential photodissociation would not skew the $^{13}$CO/C$^{18}$O ratios towards lower values.

We see higher $^{12}$CO/$^{13}$CO, $^{12}$CO/C$^{18}$O, and \\$^{13}{\rm CO/C}^{18}{\rm O}$ in the outermost, less dense regions of our targets. Although it is difficult to completely exclude preferential photodissociation in these regions, \citet{Romano2017} find the fraction of gas that resides in so-called photodissociation regions to be minimal, confirming that this mechanism is unlikely to play a key role in determining the abundance ratios in these galaxies.

\subsection{Chemical Fractionation}
The process of chemical fractionation can also drive increases in $^{13}$CO with respect to both $^{12}$CO and C$^{18}$O via the forward direction of the exchange reaction, $^{13}$C$^+ \, + \, ^{12}$CO $\rightleftharpoons \, ^{13}$CO$ \, + \, ^{12}$C$^+$ \citep{Watson1976}. This enhancement is sensitive to temperature with the forward direction dominating where gas temperatures are $\lesssim$ 20-30 K while, at higher temperatures, both directions are equally likely \citep{Smith1980,Romano2017}. Theoretical modeling of the dense molecular gas in Arp 220 and IRAS 13120-5453 suggests temperatures in local ULIRGs comfortably exceed this threshold \citep{Sakamoto1999,Sliwa2017}. In any case, increasing $^{13}$CO abundance would lead to larger, not smaller $^{13}$CO/C$^{18}$O ratios. It is therefore an unlikely pathway for producing the observed line ratios.

\subsection{Very Recent Star Formation}
\citet{Sliwa2017} \textbf{are} able to reproduce enhanced 12C/13C ratios in IRAS 13120-5453 using a Kroupa IMF \citep{Kroupa2001} if the system has undergone a very recent ($\lesssim$7 Myr) burst of star formation. A scenario whereby the extreme ratios are driven by significant, recent star formation is also invoked by \citet{Matsushita2009} in the case of Arp 220. In both cases, the recent nature of the starburst is critical. I\textbf{}f star formation is responsible for the observed deficiency in $^{13}$CO with respect to C$^{18}$O across all our targets, one would expect that $^{13}$C produced in the intermediate mass stars would begin to be ejected into the ISM on timescales of $\sim$30 Myrs \citep{Schaller1992}. Mixing between the newly produced and ejected $^{13}$C and the metals already present in the optically thin ISM would therefore drive the $^{13}$CO/C$^{18}$O line ratio from around unity and below toward the values found in more typical star-forming galaxies ($\sim$7-10).

% \citet{Sliwa2017} also find that, in the case of a recent starburst, the $^{12}$CO/$^{13}$CO value of the eject should be extremely high ($\sim$ 575). However, this a factor of twenty to fifty larger than the ratios we observe in our targets ($\sim30-100)$) which are only elevated above 'normal' values ($\sim30-60$) by a factor of two. The observed $^{13}$CO/C$^{18}$O ratios of unity or below are 7-10 times lower than those observed in normal star-forming galaxies. A recent burst of star formation does not easily explain the pattern of line ratios observed in all four targets.}

While we cannot entirely exclude recent star formation as the driving factor, the low ratios in NGC 2623, a galaxy that finished the merging process 80 Myrs ago \citep{Privon2013}, do not fit well with this mechanism. Indeed, a study of the star formation history of NGC 2623 by \citet{Cortijo-Ferrero2017} found that stars younger than 30 Myrs contribute only $\sim20\%$ of the total light density in this galaxy. Thus, we suggest it is unlikely that the observed extreme ratios in NGC 2623 are caused by a very recent ($\lesssim$7 Myr) burst of star formation.

\subsection{A Top-Heavy Initial Mass Function}
Short-lived massive stars ($\leq$30 Myr, $\gtrsim 8 {\rm M}_\odot$) are the predominant source of $^{18}$O in the ISM. In contrast, the $^{13}$C atom is dredged-up via convection into the envelopes of long-lived, low-mass stars ($\lesssim 8 {\rm M}_\odot$) that are entering the red-giant phase. These stars replenish the ISM over a longer timescale than their more massive counterparts \citep[$\sim10^9$ years;][]{Vigroux1976,Schaller1992}. Furthermore, the fact that the low $^{13}$CO/C$^{18}$O ratios are observed in all 3 ULIRGs and the post-merger system is tentative evidence that the extreme line ratios are insensitive to evolution in the merger stage.  Therefore, a stellar IMF biased toward such massive stars in our targets is a plausible explanation for the observed $^{13}$CO/C$^{18}$O ratios.

\section{Conclusions}
\label{sec:conclusions}
In this paper, we use state-of-the-art ALMA observations of molecular gas in a sample of 4 nearby merger systems to probe isotopologue emission line ratios. In doing so we find unusual ratios compared to typical star-forming galaxies and test physical mechanisms for driving this trend. Our main conclusions are as follows:

\begin{enumerate}[i)]
  \item We confirm and expand upon the results of \citet{Greve2009, Matsushita2009} and \citet{Sliwa2017}, reporting unusually low values of the isotopologue line intensity ratio $^{13}$CO/C$^{18}$O $J$=1-0 ($\lesssim1$) across the spatially resolved central regions of 3 nearby ULIRGs (Arp 220, IRAS 13120-5453, IRAS 17208-0014) and 1 nearby post-merger galaxy (NGC 2623) with respect to normal star-forming disks.
  \item We also find the isotopologue line ratios $^{12}$CO/C$^{18}$O and $^{12}$CO/$^{13}$CO $J$=1-0 are unusually large, supporting results by previous work \citep[e.g.,][]{Aalto1991,Casoli1992,Paglione2001,Sliwa2013}.
  \item We suggest that the presence of very low $^{13}$CO/C$^{18}$O line ratios in our sample and their persistence across systems at different stages of the merger process suggests C$^{18}$O is more abundant than $^{13}$CO in the molecular gas of these systems. We find this to be consistent with a scenario in which an excess in the C$^{18}$O isotopologue is driven by excess formation of massive stars ($>8 \, {\rm M}_\odot$) in these extreme environments i.e., a top-heavy stellar IMF.
\end{enumerate}

These results contribute to a growing body of evidence that the prevailing stellar IMF in nearby ULIRGs is top-heavy, and highlights the potential of submillimeter observations as probes of the IMF in galaxies that are inaccessible via traditional techniques. The stellar IMF is central to our understanding galaxy formation and, although beyond the scope of this paper, further work must conduct a full abundance analysis of local ULIRGs using state-of-the-art theoretical models that account for optical depth effects. Such a study would make a considerable contribution to our knowledge of starburst galaxies, both locally and at high redshift.

\acknowledgments
We would like to thank the referee for their detailed and constructive comments. T.\ Brown also thanks A.\ Leroy for helpful discussions that certainly improved this paper. This paper makes use of the following ALMA data: ADS/JAO.ALMA\#2013.1.00379.S, ADS/JAO.ALMA\#2015.1.00167.S, ADS/JAO.ALMA\#\\2015.1.00287.S, ADS/JAO.ALMA\#2015.1.01191.S, \\ADS/JAO.ALMA\#2016.1.00140.S, ADS/JAO.ALMA\\\#2017.1.01306.S. \\ALMA is a partnership of ESO (representing its member states), NSF (USA) and NINS (Japan), together with NRC (Canada), MOST and ASIAA (Taiwan), and KASI (Republic of Korea), in cooperation with the Republic of Chile. The Joint ALMA Observatory is operated by ESO, AUI/NRAO and NAOJ. The National Radio Astronomy Observatory is a facility of the National Science Foundation operated under cooperative agreement by Associated Universities, Inc. This research has made use of the NASA/IPAC Extragalactic Database (NED) which is operated by the Jet Propulsion Laboratory, California Institute of Technology, under contract with the National Aeronautics and Space Administration. CDW acknowledges financial support from the Canada Council for the Arts through a Killam Research Fellowship. The research of CDW is supported by grants from the Natural Sciences and Engineering Research Council of Canada and the Canada Research Chairs program. 

% \bibliography{refs}

\begin{thebibliography}

\bibitem[Aalto et al.(1991)]{Aalto1991} Aalto, S., Black, J.~H., Johansson, L.~E.~B., et al.\ 1991, \aap, 249, 323.
\bibitem[Aalto et al.(2015)]{Aalto2015} Aalto, S., Mart{\'\i}n, S., Costagliola, F., et al.\ 2015, \aap, 584, A42.
\bibitem[Armus et al.(1989)]{Armus1989} Armus, L., Heckman, T.~M., \& Miley, G.~K.\ 1989, \apj, 347, 727.
\bibitem[Armus et al.(2009)]{Armus2009} Armus, L., Mazzarella, J.~M., Evans, A.~S., et al.\ 2009, Publications of the Astronomical Society of the Pacific, 121, 559.
% \bibitem[Bally \& Langer(1982)]{Bally1982} Bally, J., \& Langer, W.~D.\ 1982, \apj, 255, 143.
\bibitem[Barnes \& Hernquist(1991)]{Barnes1991} Barnes, J.~E., \& Hernquist, L.~E.\ 1991, \apj, 370, L65.
\bibitem[Barnes \& Hernquist(1992)]{Barnes1992} Barnes, J.~E., \& Hernquist, L.\ 1992, Annual Review of Astronomy and Astrophysics, 30, 705.
\bibitem[Bastian et al.(2010)]{Bastian2010} Bastian, N., Covey, K.~R., \& Meyer, M.~R.\ 2010, Annual Review of Astronomy and Astrophysics, 48, 339.
\bibitem[Briggs(1995)]{Briggs1995} Briggs, D.~S.\ 1995, American Astronomical Society Meeting Abstracts 187, 112.02.
\bibitem[Casoli et al.(1988)]{Casoli1988} Casoli, F., Combes, F., Dupraz, C., et al.\ 1988, \aap, 192, L17.
\bibitem[Casoli et al.(1992)]{Casoli1992} Casoli, F., Dupraz, C., \& Combes, F.\ 1992, \aap, 264, 55.
\bibitem[Chabrier(2003)]{Chabrier2003} Chabrier, G.\ 2003, Publications of the Astronomical Society of the Pacific, 115, 763.
\bibitem[Chou et al.(2007)]{Chou2007} Chou, R.~C.~Y., Peck, A.~B., Lim, J., et al.\ 2007, \apj, 670, 116.
\bibitem[Cortijo-Ferrero et al.(2017)]{Cortijo-Ferrero2017} Cortijo-Ferrero, C., Gonz{\'a}lez Delgado, R.~M., P{\'e}rez, E., et al.\ 2017, \aap, 607, A70.
\bibitem[Danielson et al.(2013)]{Danielson2013} Danielson, A.~L.~R., Swinbank, A.~M., Smail, I., et al.\ 2013, \mnras, 436, 2793.
\bibitem[Falstad et al.(2015)]{Falstad2015} Falstad, N., Gonz{\'a}lez-Alfonso, E., Aalto, S., et al.\ 2015, \aap, 580, A52.
\bibitem[Genzel et al.(1998)]{Genzel1998} Genzel, R., Lutz, D., Sturm, E., et al.\ 1998, \apj, 498, 579.
\bibitem[Gonz{\'a}lez \& Koenigsberger(2014)]{Gonzalez2014} Gonz{\'a}lez, R.~F., \& Koenigsberger, G.\ 2014, \aap, 561, A105.
\bibitem[Greve et al.(2009)]{Greve2009} Greve, T.~R., Papadopoulos, P.~P., Gao, Y., et al.\ 2009, \apj, 692, 1432.
\bibitem[Henkel \& Mauersberger(1993)]{Henkel1993} Henkel, C., \& Mauersberger, R.\ 1993, \aap, 274, 730.
\bibitem[Iben(1975)]{Iben1975} Iben, I.\ 1975, \apj, 196, 525.
\bibitem[Jim{\'e}nez-Donaire et al.(2017)]{Jimenez-Donaire2017} Jim{\'e}nez-Donaire, M.~J., Cormier, D., Bigiel, F., et al.\ 2017, \apj, 836, L29.
\bibitem[K{\"o}nig et al.(2016)]{Konig2016} K{\"o}nig, S., Aalto, S., Muller, S., et al.\ 2016, \aap, 594, A70.
\bibitem[Kroupa(2001)]{Kroupa2001} Kroupa, P.\ 2001, \mnras, 322, 231.
\bibitem[Langer \& Penzias(1993)]{Langer1993} Langer, W.~D., \& Penzias, A.~A.\ 1993, \apj, 408, 539.
\bibitem[Lonsdale et al.(2006)]{Lonsdale2006} Lonsdale, C.~J., Farrah, D., \& Smith, H.~E.\ 2006, Astrophysics Update 2, 285.
\bibitem[Madau \& Dickinson(2014)]{Madau2014} Madau, P., \& Dickinson, M.\ 2014, Annual Review of Astronomy and Astrophysics, 52, 415.
\bibitem[Mart{\'{\i}}n et al.(2019)]{Martin2019} Mart{\'{\i}}n, S., Muller, S., Henkel, C., et al.\ 2019, \aap, 624, A125 
\bibitem[Matsushita et al.(2009)]{Matsushita2009} Matsushita, S., Iono, D., Petitpas, G.~R., et al.\ 2009, \apj, 693, 56.
\bibitem[Matteucci(2012)]{Matteucci2012} Matteucci, F.\ 2012, Chemical Evolution of Galaxies:.
\bibitem[McMullin et al.(2007)]{McMullin2007} McMullin, J.~P., Waters, B., Schiebel, D., et al.\ 2007, Astronomical Data Analysis Software and Systems XVI, 127.
\bibitem[Meier \& Turner(2004)]{Meier2004} Meier, D.~S., \& Turner, J.~L.\ 2004, \aj, 127, 2069.
\bibitem[Meynet \& Maeder(2002)]{Meynet2002} Meynet, G., \& Maeder, A.\ 2002, \aap, 390, 561.
% \bibitem[Milam et al.(2005)]{Milam2005} Milam, S.~N., Savage, C., Brewster, M.~A., et al.\ 2005, \apj, 634, 1126.
\bibitem[Mirabel \& Sanders(1989)]{Mirabel1989} Mirabel, I.~F., \& Sanders, D.~B.\ 1989, \apj, 340, L53.
\bibitem[Murphy et al.(1999)]{Murphy1999} Murphy, T.~W., Soifer, B.~T., Matthews, K., et al.\ 1999, \apj, 525, L85.
\bibitem[Papadopoulos et al.(1996)]{Papadopoulos1996} Papadopoulos, P.~P., Seaquist, E.~R., \& Scoville, N.~Z.\ 1996, \apj, 465, 173.
\bibitem[Privon et al.(2013)]{Privon2013} Privon, G.~C., Barnes, J.~E., Evans, A.~S., et al.\ 2013, \apj, 771, 120.
\bibitem[Privon et al.(2017)]{Privon2017} Privon, G.~C., Aalto, S., Falstad, N., et al.\ 2017, \apj, 835, 213.
\bibitem[Paglione et al.(2001)]{Paglione2001} Paglione, T.~A.~D., Wall, W.~F., Young, J.~S., et al.\ 2001, The Astrophysical Journal Supplement Series, 135, 183.
\bibitem[Romano et al.(2017)]{Romano2017} Romano, D., Matteucci, F., Zhang, Z.-Y., et al.\ 2017, \mnras, 470, 401.
% \bibitem[Rupke et al.(2008)]{Rupke2008} Rupke, D.~S.~N., Veilleux, S., \& Baker, A.~J.\ 2008, \apj, 674, 172.
\bibitem[Sage et al.(1991)]{Sage1991} Sage, L.~J., Mauersberger, R., \& Henkel, C.\ 1991, \aap, 249, 31.
\bibitem[Schaller et al.(1992)]{Schaller1992} Schaller, G., Schaerer, D., Meynet, G., et al.\ 1992, Astronomy and Astrophysics Supplement Series, 96, 269.
\bibitem[Scoville et al.(2017)]{Scoville2017} Scoville, N., Murchikova, L., Walter, F., et al.\ 2017, \apj, 836, 66.
\bibitem[Sliwa et al.(2017)]{Sliwa2017} Sliwa, K., Wilson, C.~D., Aalto, S., et al.\ 2017, \apj, 840, L11.
\bibitem[Sliwa et al.(2013)]{Sliwa2013} Sliwa, K., Wilson, C.~D., Krips, M., et al.\ 2013, \apj, 777, 126.
\bibitem[Sakamoto et al.(1999)]{Sakamoto1999} Sakamoto, K., Scoville, N.~Z., Yun, M.~S., et al.\ 1999, \apj, 514, 68.
\bibitem[Sanders \& Mirabel(1996)]{Sanders1996} Sanders, D.~B., \& Mirabel, I.~F.\ 1996,  Annual Review of Astronomy and Astrophysics, 34, 749.
\bibitem[Smith \& Adams(1980)]{Smith1980} Smith, D., \& Adams, N.~G.\ 1980, \apj, 242, 424.
\bibitem[van Dishoeck \& Black(1988)]{vanDishoeck1988} van Dishoeck, E.~F., \& Black, J.~H.\ 1988, \apj, 334, 771.
\bibitem[Vigroux et al.(1976)]{Vigroux1976} Vigroux, L., Audouze, J., \& Lequeux, J.\ 1976, \aap, 52, 1.
\bibitem[Watson et al.(1976)]{Watson1976} Watson, W.~D., Anicich, V.~G., \& Huntress, W.~T.\ 1976, \apj, 205, L165.
\bibitem[Wouterloot et al.(2008)]{Wouterloot2008} Wouterloot, J.~G.~A., Henkel, C., Brand, J., et al.\ 2008, \aap, 487, 237.
\bibitem[Wilson \& Matteucci(1992)]{Wilson1992} Wilson, T.~L., \& Matteucci, F.\ 1992, Astronomy and Astrophysics Review, 4, 1.
% \bibitem[Wilson \& Rood(1994)]{Wilson1994} Wilson, T.~L., \& Rood, R.\ 1994, Annual Review of Astronomy and Astrophysics, 32, 191.
\bibitem[Zhang et al.(2018)]{Zhang2018} Zhang, Z.-Y., Romano, D., Ivison, R.~J., et al.\ 2018, \nat, 558, 260.

\end{thebibliography}

\appendix{}
\setcounter{figure}{0} \renewcommand{\thefigure}{A.\arabic{figure}} Figure \ref{fig:spectra} shows the mean integrated flux averaged over each target. Fluxes are extracted from final data cubes (cleaned and smoothed to 35 km/s) using an aperture that contains all the detected C$^{18}$O and $^{13}$CO line emission.

\begin{figure}
\centering
\includegraphics[width=\textwidth]{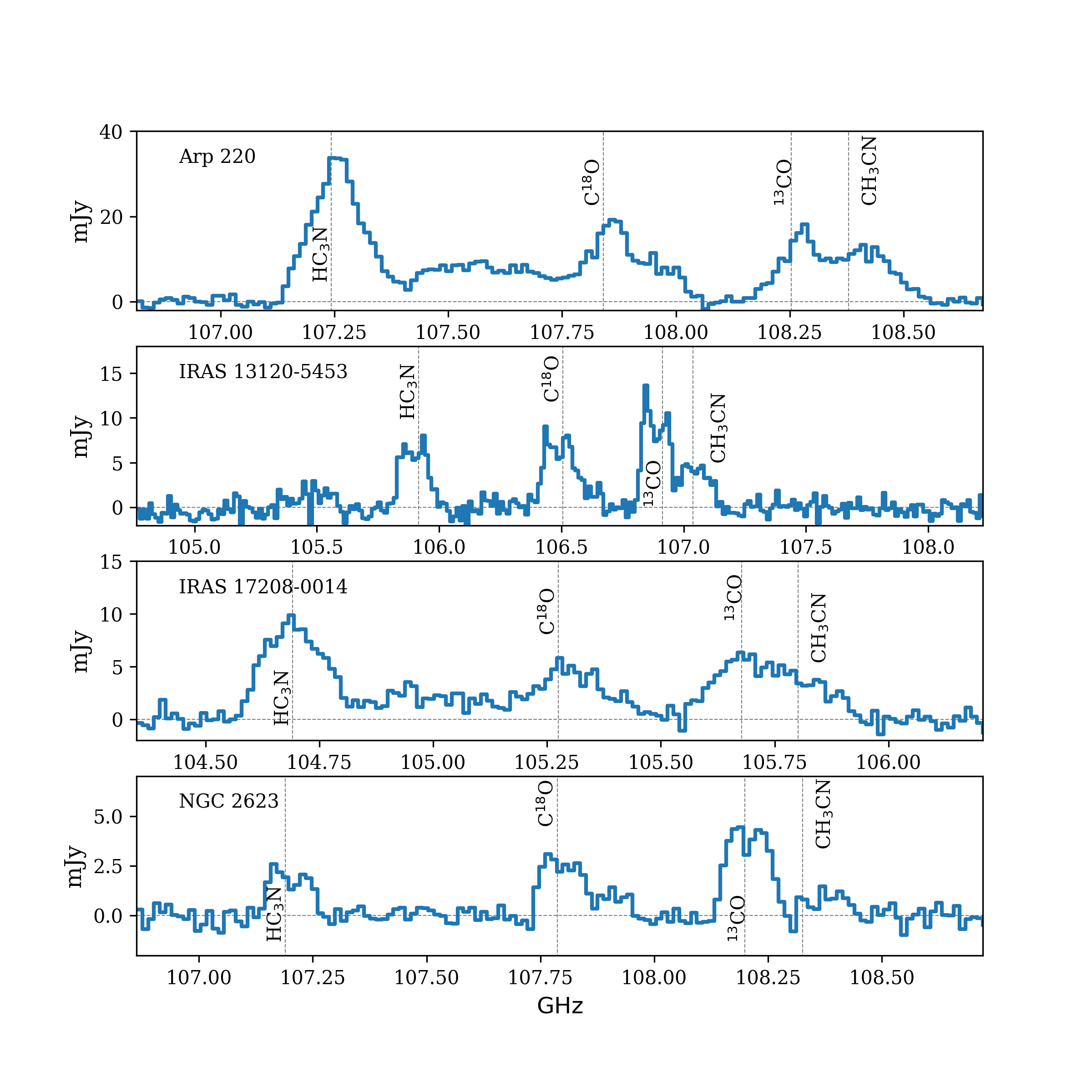}
\caption{Spectra for each target galaxy showing the C$^{18}$O and $^{13}$CO $J=1-0$ emission lines. The other emission lines present are ${\rm HC_3N}$ and ${\rm CH_3CN}$.}
\label{fig:spectra}
\end{figure}



\end{document}